\documentclass[aps, prl,superscriptaddress,twocolumn,secnumarabic,amssymb, nobibnotes,floatfix]{revtex4-2}
\usepackage{graphicx}
\usepackage{dcolumn}

\begin{document}

\title{Search for Relativistic Magnetic Monopoles with Eight Years of IceCube Data}%

\affiliation{III. Physikalisches Institut, RWTH Aachen University, D-52056 Aachen, Germany}
\affiliation{Department of Physics, University of Adelaide, Adelaide, 5005, Australia}
\affiliation{Department of Physics and Astronomy, University of Alaska Anchorage, 3211 Providence Dr., Anchorage, AK 99508, USA}
\affiliation{Department of Physics, University of Texas at Arlington, 502 Yates St., Science Hall Rm 108, Box 19059, Arlington, TX 76019, USA}
\affiliation{CTSPS, Clark-Atlanta University, Atlanta, GA 30314, USA}
\affiliation{School of Physics and Center for Relativistic Astrophysics, Georgia Institute of Technology, Atlanta, GA 30332, USA}
\affiliation{Department of Physics, Southern University, Baton Rouge, LA 70813, USA}
\affiliation{Department of Physics, University of California, Berkeley, CA 94720, USA}
\affiliation{Lawrence Berkeley National Laboratory, Berkeley, CA 94720, USA}
\affiliation{Institut f{\"u}r Physik, Humboldt-Universit{\"a}t zu Berlin, D-12489 Berlin, Germany}
\affiliation{Fakult{\"a}t f{\"u}r Physik {\&} Astronomie, Ruhr-Universit{\"a}t Bochum, D-44780 Bochum, Germany}
\affiliation{Universit{\'e} Libre de Bruxelles, Science Faculty CP230, B-1050 Brussels, Belgium}
\affiliation{Vrije Universiteit Brussel (VUB), Dienst ELEM, B-1050 Brussels, Belgium}
\affiliation{Department of Physics and Laboratory for Particle Physics and Cosmology, Harvard University, Cambridge, MA 02138, USA}
\affiliation{Department of Physics, Massachusetts Institute of Technology, Cambridge, MA 02139, USA}
\affiliation{Department of Physics and Institute for Global Prominent Research, Chiba University, Chiba 263-8522, Japan}
\affiliation{Department of Physics, Loyola University Chicago, Chicago, IL 60660, USA}
\affiliation{Department of Physics and Astronomy, University of Canterbury, Private Bag 4800, Christchurch, New Zealand}
\affiliation{Department of Physics, University of Maryland, College Park, MD 20742, USA}
\affiliation{Department of Astronomy, Ohio State University, Columbus, OH 43210, USA}
\affiliation{Department of Physics and Center for Cosmology and Astro-Particle Physics, Ohio State University, Columbus, OH 43210, USA}
\affiliation{Niels Bohr Institute, University of Copenhagen, DK-2100 Copenhagen, Denmark}
\affiliation{Department of Physics, TU Dortmund University, D-44221 Dortmund, Germany}
\affiliation{Department of Physics and Astronomy, Michigan State University, East Lansing, MI 48824, USA}
\affiliation{Department of Physics, University of Alberta, Edmonton, Alberta, Canada T6G 2E1}
\affiliation{Erlangen Centre for Astroparticle Physics, Friedrich-Alexander-Universit{\"a}t Erlangen-N{\"u}rnberg, D-91058 Erlangen, Germany}
\affiliation{Physik-department, Technische Universit{\"a}t M{\"u}nchen, D-85748 Garching, Germany}
\affiliation{D{\'e}partement de physique nucl{\'e}aire et corpusculaire, Universit{\'e} de Gen{\`e}ve, CH-1211 Gen{\`e}ve, Switzerland}
\affiliation{Department of Physics and Astronomy, University of Gent, B-9000 Gent, Belgium}
\affiliation{Department of Physics and Astronomy, University of California, Irvine, CA 92697, USA}
\affiliation{Karlsruhe Institute of Technology, Institute for Astroparticle Physics, D-76021 Karlsruhe, Germany }
\affiliation{Karlsruhe Institute of Technology, Institute of Experimental Particle Physics, D-76021 Karlsruhe, Germany }
\affiliation{Department of Physics, Engineering Physics, and Astronomy, Queen's University, Kingston, ON K7L 3N6, Canada}
\affiliation{Department of Physics and Astronomy, University of Kansas, Lawrence, KS 66045, USA}
\affiliation{Department of Physics and Astronomy, UCLA, Los Angeles, CA 90095, USA}
\affiliation{Department of Physics, Mercer University, Macon, GA 31207-0001, USA}
\affiliation{Department of Astronomy, University of Wisconsin{\textendash}Madison, Madison, WI 53706, USA}
\affiliation{Department of Physics and Wisconsin IceCube Particle Astrophysics Center, University of Wisconsin{\textendash}Madison, Madison, WI 53706, USA}
\affiliation{Institute of Physics, University of Mainz, Staudinger Weg 7, D-55099 Mainz, Germany}
\affiliation{Department of Physics, Marquette University, Milwaukee, WI, 53201, USA}
\affiliation{Institut f{\"u}r Kernphysik, Westf{\"a}lische Wilhelms-Universit{\"a}t M{\"u}nster, D-48149 M{\"u}nster, Germany}
\affiliation{Bartol Research Institute and Department of Physics and Astronomy, University of Delaware, Newark, DE 19716, USA}
\affiliation{Department of Physics, Yale University, New Haven, CT 06520, USA}
\affiliation{Department of Physics, University of Oxford, Parks Road, Oxford OX1 3PU, UK}
\affiliation{Department of Physics, Drexel University, 3141 Chestnut Street, Philadelphia, PA 19104, USA}
\affiliation{Physics Department, South Dakota School of Mines and Technology, Rapid City, SD 57701, USA}
\affiliation{Department of Physics, University of Wisconsin, River Falls, WI 54022, USA}
\affiliation{Department of Physics and Astronomy, University of Rochester, Rochester, NY 14627, USA}
\affiliation{Department of Physics and Astronomy, University of Utah, Salt Lake City, UT 84112, USA}
\affiliation{Oskar Klein Centre and Department of Physics, Stockholm University, SE-10691 Stockholm, Sweden}
\affiliation{Department of Physics and Astronomy, Stony Brook University, Stony Brook, NY 11794-3800, USA}
\affiliation{Department of Physics, Sungkyunkwan University, Suwon 16419, Korea}
\affiliation{Institute of Basic Science, Sungkyunkwan University, Suwon 16419, Korea}
\affiliation{Department of Physics and Astronomy, University of Alabama, Tuscaloosa, AL 35487, USA}
\affiliation{Department of Astronomy and Astrophysics, Pennsylvania State University, University Park, PA 16802, USA}
\affiliation{Department of Physics, Pennsylvania State University, University Park, PA 16802, USA}
\affiliation{Department of Physics and Astronomy, Uppsala University, Box 516, S-75120 Uppsala, Sweden}
\affiliation{Department of Physics, University of Wuppertal, D-42119 Wuppertal, Germany}
\affiliation{DESY, D-15738 Zeuthen, Germany}

\author{R. Abbasi}
\affiliation{Department of Physics, Loyola University Chicago, Chicago, IL 60660, USA}
\author{M. Ackermann}
\affiliation{DESY, D-15738 Zeuthen, Germany}
\author{J. Adams}
\affiliation{Department of Physics and Astronomy, University of Canterbury, Private Bag 4800, Christchurch, New Zealand}
\author{J. A. Aguilar}
\affiliation{Universit{\'e} Libre de Bruxelles, Science Faculty CP230, B-1050 Brussels, Belgium}
\author{M. Ahlers}
\affiliation{Niels Bohr Institute, University of Copenhagen, DK-2100 Copenhagen, Denmark}
\author{M. Ahrens}
\affiliation{Oskar Klein Centre and Department of Physics, Stockholm University, SE-10691 Stockholm, Sweden}
\author{C. Alispach}
\affiliation{D{\'e}partement de physique nucl{\'e}aire et corpusculaire, Universit{\'e} de Gen{\`e}ve, CH-1211 Gen{\`e}ve, Switzerland}
\author{A. A. Alves Jr.}
\affiliation{Karlsruhe Institute of Technology, Institute for Astroparticle Physics, D-76021 Karlsruhe, Germany }
\author{N. M. Amin}
\affiliation{Bartol Research Institute and Department of Physics and Astronomy, University of Delaware, Newark, DE 19716, USA}
\author{R. An}
\affiliation{Department of Physics and Laboratory for Particle Physics and Cosmology, Harvard University, Cambridge, MA 02138, USA}
% ...

\date{\today}

\collaboration{IceCube Collaboration}
\noaffiliation


\date{January 20, 2022}%
\begin{abstract}  
  We present an all-sky 90\% confidence level upper limit on the cosmic flux of relativistic magnetic monopoles using 2886 days of IceCube data.  The analysis was optimized for monopole speeds between 0.750$c$ and 0.995$c$, without any explicit restriction on the monopole mass. We constrain the flux of relativistic cosmic magnetic monopoles to a level below  $2.0\times 10^{-19} {\textrm{cm}}^{-2} {\textrm{s}}^{-1} {\textrm{sr}}^{-1}$ over the majority of the targeted speed range. This result constitutes the most strict upper limit to date for magnetic monopoles with $\beta \gtrsim 0.8$ and up to $\beta \sim 0.995$ and fills the gap between existing limits on the cosmic flux of nonrelativistic and ultrarelativistic magnetic monopoles.
\end{abstract}
\maketitle

\paragraph{Introduction.---} Although not yet experimentally detected, particles with magnetic charge (magnetic monopoles) are explicitly allowed in theories describing the fundamental laws of physics. Maxwell's equations allow for the introduction of a magnetic current and a magnetic charge density without loss of internal consistency or contradiction with experimental results. Magnetic monopoles were introduced in quantum mechanics by Dirac in 1931~\cite{Dirac:1931kp} through a mechanism that requires the quantization of both the electric and magnetic charge. The allowed magnetic charges are given by
$g=Ne/(2\alpha)$, where $\alpha$ is the fine structure constant, $e$ is the unit electric charge and $N$ is an integer. The smallest allowed magnetic charge is thus $g=e/2\alpha=68.5e$, also called the Dirac charge, $g\strut_{\mathrm D}$. Magnetic monopoles can also be accommodated in grand unified theories (GUTs) of the known forces in nature, which are based on quantum field theory. GUTs predict the existence of stable magnetic monopoles ('t Hooft-Polyakov monopoles), that arise when the gauge symmetry of the GUT breaks into the electromagnetic U(1) symmetry, creating a topological soliton that must carry magnetic charge~\cite{tHooft:1974kcl,Polyakov:1974ek,Goddard:1978aaa}. The mass of a monopole depends on the details of the symmetry breaking from the GUT symmetry to U(1). GUT monopoles have masses on the order of $10^{17}$~GeV, but so-called intermediate mass monopoles with masses $\gtrsim 10^5$~GeV can arise if there is a mass scale between the GUT scale and the electroweak scale~\cite{Zyla:2020zbs}.\par
Magnetic monopoles could have formed in the early Universe as the temperature of the primordial plasma dropped below the energy scale of the GUT symmetry breaking~\cite{Kibble:1976sj,Preskill:1979zi}. The expected production rate of monopoles depends on the unknown nature of this phase transition (first or second order), but it can lead to a production comparable to the amount of baryons, predicting a relic density today above current observational limits. This has been dubbed the ``monopole problem''~\cite{Preskill:1984gd}. It is through inflation that the primordial density of magnetic monopoles can be brought to a level consistent with observations. Remaining magnetic monopoles will be accelerated along the magnetic field lines of galactic and extragalactic magnetic fields, gaining a kinetic energy given by $E_{\mathrm{k}}=g\strut_{\mathrm D}|\vec{B}|L$, where $|\vec{B}|$ is the strength of the magnetic field and $L$ represents the size of the domain where the magnetic field direction remains constant. Even if typical intergalactic and galactic magnetic fields are weak (order of nano- to microGauss)~\cite{Han:2017aaa}, their spatial extensions are large (kilo- to Megaparsec), so the total energy gain can be substantial for a monopole crossing several magnetic field domains over the lifetime of the Universe~\cite{Wick:2000yc,Spurio:2019oaq}. Indeed energies up to $\sim 10^{14}$~GeV can be expected. Depending on the mass of the monopole, this can result in relativistic speeds in the present epoch. The expected speed range of relic monopoles is an important aspect for their detection, since different techniques are used to search for monopoles of different speeds~\cite{Wick:2000yc,Patrizii:2019eud,Spurio:2019oaq}. 

Relativistic monopoles (defined as those with $\beta \geq 0.750$ in what follows) will induce Cherenkov radiation when traversing a dielectric medium faster than the speed of light in the medium, as electrically charged particles do~\cite{Tompkins:1965aaa}. Therefore Cherenkov detectors can be used to search for cosmic monopoles through the pattern they would leave when passing through the detector~\cite{Aartsen:2014awd,Aartsen:2015exf,Albert:2017fud,Antipin:2007zz}. The analysis presented in this Letter uses eight years of data from the IceCube neutrino observatory to search for monopoles with one unit of magnetic charge and speed above 0.750$c$.

IceCube is a cubic-kilometer array of digital optical modules (DOMs) deployed in 86 strings at depths between 1.45 and 2.45~km below the surface of the glacial ice at the South Pole~\cite{Aartsen:2016nxy}. IceCube uses the ice both as target and detection medium to detect the Cherenkov light from secondary particles produced in neutrino interactions, but it is also sensitive to searches for any new particle that can produce light that is detectable by the optical modules. An air shower array on the surface, IceTop, is used for cosmic ray studies and also serves as a veto for atmospheric-induced events in IceCube~\cite{IceCube:2012nn}.

\paragraph{Signal.---} The equivalent of the Frank-Tamm formula for the amount of Cherenkov light emitted by a relativistic monopole shows that it produces $(g n /e)^2$ times as much Cherenkov light as a particle with electric charge $e$ and the same speed in a medium with refractive index $n$~\cite{Tompkins:1965aaa}. In ice ($n=1.33$, slightly wavelength dependent~\cite{Price:2000yh}), a single--charge monopole above the Cherenkov threshold ($\beta=0.750$) emits about 8300 times more photons per unit length than a minimum ionizing muon. Additionally, indirect Cherenkov light from the products of ionization is a non-negligible contribution, mainly close to the Cherenkov threshold. The signature of a magnetic monopole crossing IceCube is therefore expected to be a straight track with uniform light yield along its path. This will be a distinctive selection criterion in the analysis presented below.  For ultrarelativistic monopoles ($\gamma \gtrsim 10^3$),  secondary particles produced in stochastic energy losses like pair production, bremsstrahlung and photonuclear interactions~\cite{Ahlen:1978jy,Derkaoui:1998uv} are an important and non-negligible source of additional Cherenkov light. The analysis of such speeds is beyond the scope of this work.

A sample of 4$\times$10$^5$ through-going monopole events was simulated with the standard IceCube simulation software, which includes light production and propagation, as well as the detector trigger response. A uniform speed distribution between 0.750$c$ and 0.995$c$ at the detector was adopted, and isotropic arrival directions were assumed.  Relativistic monopoles can lose energy when traversing matter and they might become nonrelativistic, or even stop, before reaching the detector. The energy loss of a 1$g\strut_{\mathrm{ D}}$ relativistic monopole in rock(iron) is $\sim 10(100)$~GeV/cm~\cite{Derkaoui:1998uv}. A vertically up-going monopole crossing one Earth diameter ($10^{9}$~cm) will lose about $10^{11}$~GeV, while a monopole reaching the detector horizontally will just cross a chord of $\sim 10^7$~cm and lose $10^{8}$~GeV. This determines a lower mass limit for the validity of the analysis of $m_{\mathrm{M}}\sim 10^{8-11}$~GeV, depending on the zenith angle, to ensure that monopoles reach the detector.

 \paragraph{Backgrounds.---} The main background in the search for relativistic monopoles with IceCube are very energetic astrophysical neutrinos~\cite{Aartsen:2015knd,Aartsen:2016xlq}. Muons from astrophysical neutrino interactions produce enough light in the detector to mimic monopole tracks. Their distinctive feature is the stochastic emission of light along the muon track.  The expected number of astrophysical background events at different levels of the analysis has been calculated from the measured  astrophysical neutrino flux~\cite{Haack:2017dxi} which can be described by a single power law, $\Phi_{\nu}=\Phi_{\mathrm 0}\times (E_{\nu}/100~{\mathrm TeV})^{\gamma}$, with $\Phi_{\mathrm 0}=1.01^{+0.26}_{-0.23}\times 10^{-18}\mathrm{GeV^{-1}}\mathrm{cm^{-2}}\mathrm{s^{-1}}\mathrm{sr^{-1}}$, and $\gamma=2.19\pm 0.10$. We assume no cutoff in the spectrum and a $\nu_e:\nu_\mu:\nu_\tau$ flavor ratio at Earth of 1:1:1.

Other backgrounds to this analysis are atmospheric neutrinos and the much more copious atmospheric muons produced in cosmic ray interactions in the atmosphere and that reach the detector depth from above. As explained below, these backgrounds are removed by either a directional selection (considering only up-going events removes atmospheric muons) or by the event characteristics (muons from atmospheric neutrinos produce relatively dim muon tracks that do not resemble the much brighter signature of a monopole). \\

\begin{figure*}[t]
\begin{minipage}{0.48\linewidth}
    \includegraphics[width=\textwidth]{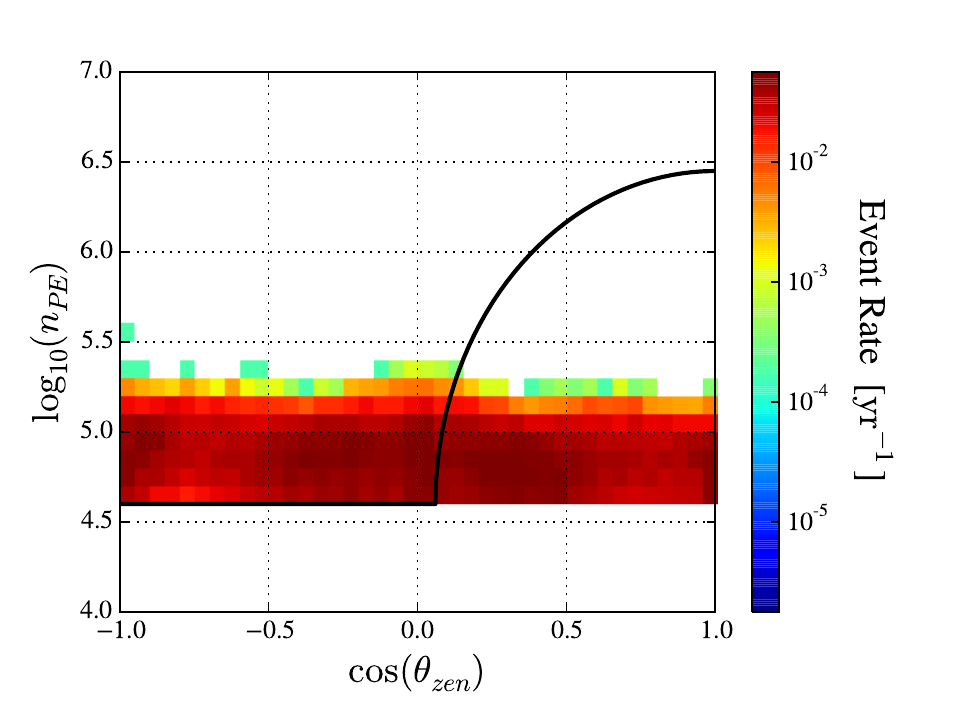}
\end{minipage}
\begin{minipage}{0.48\linewidth}
    \includegraphics[width=\textwidth]{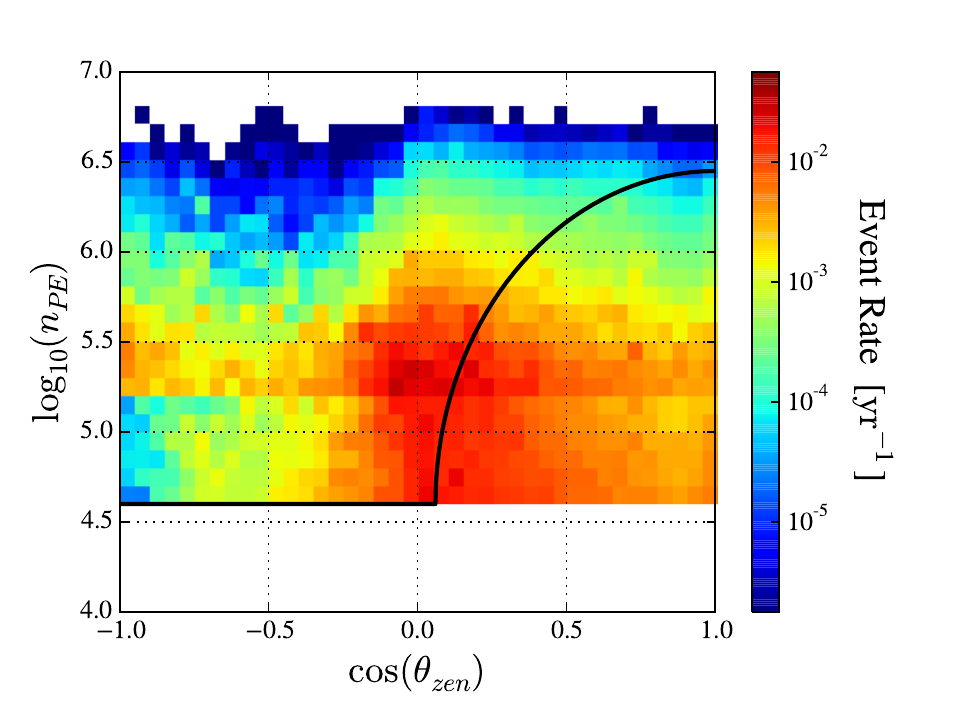}
\end{minipage}
\caption{(Left) Expected event rate of monopoles as a function of number of photoelectrons ($n_{\mathrm{PE}}$) and zenith angle of the track reconstruction ($\theta_{\mathrm{zen}}$). The color scale is normalized to the expected monopole event rate assuming a flux equal to the previous IceCube limit~\cite{Aartsen:2015exf}. (Right) Same for astrophysical neutrinos, with the color scale representing the event rate normalized to the measured flux. $\mathrm{cos}({\theta_{\mathrm{zen}}})=-1$ corresponds to vertically up-going tracks while $\mathrm{cos}({\theta_{\mathrm{zen}}})=1$ to vertically down-going tracks. Events with less than $10^{4.6}$ $n_{\mathrm{PE}}$ have been rejected by an early cut in step I, while the black line shows the effect of the final, angular-dependent, step I selection (see the text for details). Events below the black line are discarded by this selection.}
    \label{fig:stepI_cut}
\end{figure*}

\paragraph{Data selection.---} This analysis uses IceCube data collected between 2011 and 2018, corresponding to 2886~days of live time~\cite{Burgman:2021aaa}. The event selection was performed in two steps using all the IceCube triggered events. The first selection (denoted as step I in the rest of the Letter) consists of applying the exact criteria (``cuts'') that were developed for a previous IceCube analysis searching for extremely high-energy neutrinos, and described in detail in~\cite{Aartsen:2018vtx}. Detailed comparisons between data and Monte Carlo simulations were performed in that analysis to validate the simulation used to generate the different backgrounds and evaluate the efficiency of the event selection.\par

The analysis strategy of step I aims at selecting very bright events, potentially corresponding to extremely high energetic astrophysical neutrinos, and is a good base to further develop a selection for relativistic monopole events, which would also leave a bright event pattern in the detector. An online event filter at the South Pole selects events with more than 1000 photoelectrons  ($n_{\mathrm{PE}}$) to reduce the amount of data transferred by satellite for further processing in the northern hemisphere. Additional aggressive brightness cuts, selecting events with a high number of $n_{\mathrm{PE}}$, are applied off-line in order to drastically reduce the number of atmospheric background events. 
At least $10^{4.6}$ photoelectrons are required for events with a good track fit quality, increasing linearly with the reduced log-likelihood of the fit to $n_{\mathrm{PE}} \geq 10^{5.2}$ for events with lower track reconstruction quality. This cut has been shown to select muon tracks relative to cascade like events produced by electron  and tau neutrino interactions. 
Additionally, a zenith-angle dependent $n_{\mathrm{PE}}$ cut is applied to remove high energetic atmospheric muons that reach the detector from above the horizon. The black line of Fig.~\ref{fig:stepI_cut} illustrates this selection. Down-going events registered in coincidence with two or more pulses in the IceTop surface air shower array are also rejected as probably induced by an atmospheric shower. 
One of the effects of the step I cuts is to reject down-going monopoles, resulting in uniform acceptance of the analysis in $\mathrm{cos}({\theta_{\mathrm{zen}}})$ for monopoles reaching the detector from below the horizon and no acceptance for monopoles coming from above, see the left plot of Fig.~\ref{fig:stepI_cut}. Table~\ref{tab:results} shows the remaining fraction of signal and background events after the step I cut level. \par

The step I event selection has practically no acceptance for neutrinos with energy below $10^{6}$~GeV, where the bulk of atmospheric neutrinos lie. Atmospheric muons and neutrinos are reduced to a rate of about $10^{-10}$~Hz after the step I brightness and zenith angle selection, which is well below 0.1 event in the whole analysis live time. We therefore assume a zero background from atmospheric muons and atmospheric neutrinos in what follows.\par

\begin{figure*}[t]
\begin{minipage}{0.48\linewidth}
    \includegraphics[width=\textwidth]{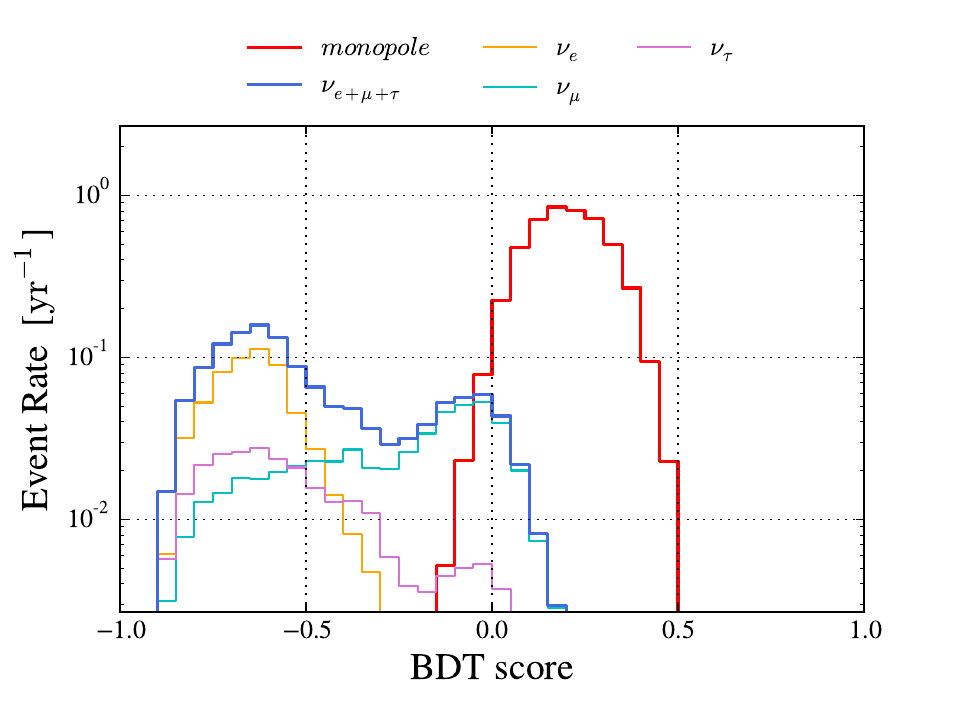}
\end{minipage}
\begin{minipage}{0.48\linewidth}
    \includegraphics[width=\textwidth]{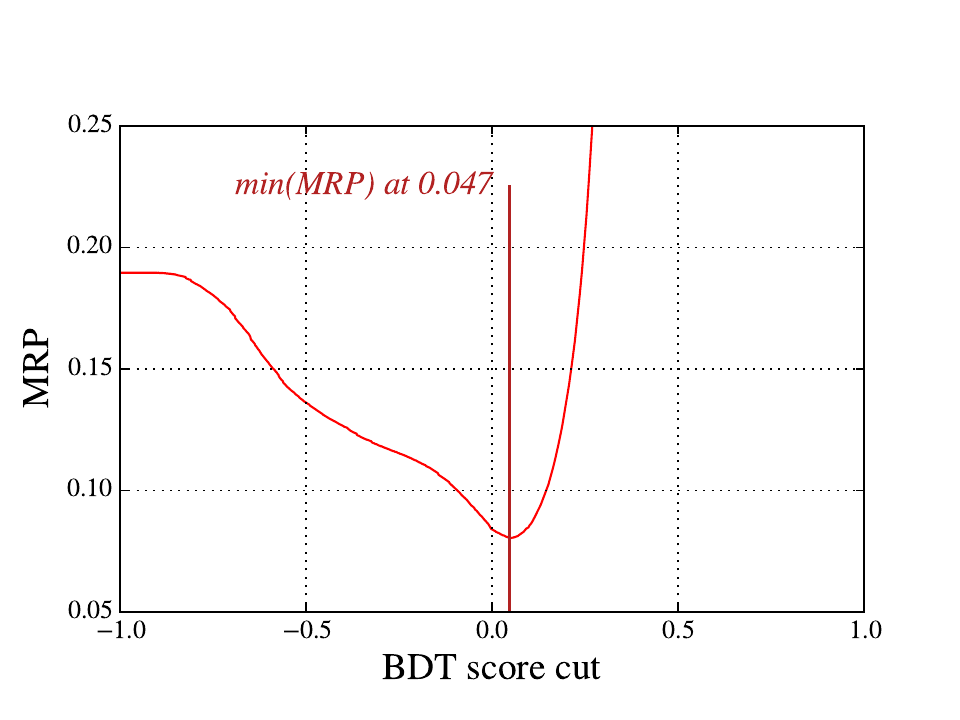}
\end{minipage}
\caption{(Left) BDT score for the signal (monopoles, normalized to the previous IceCube limit~\cite{Aartsen:2015exf}) and the different background components (astrophysical electron, muon and tau neutrinos, as well as their sum, normalized according to the measured flux~\cite{Haack:2017dxi}). (Right) The model rejection potential as a function of BDT score, showing a clear minimum at 0.047 that defines the cut value that provides the best sensitivity of the analysis.}
    \label{fig:BDT_score}
\end{figure*}

\begin{figure}
    \centering
    \includegraphics[width=0.5\textwidth]{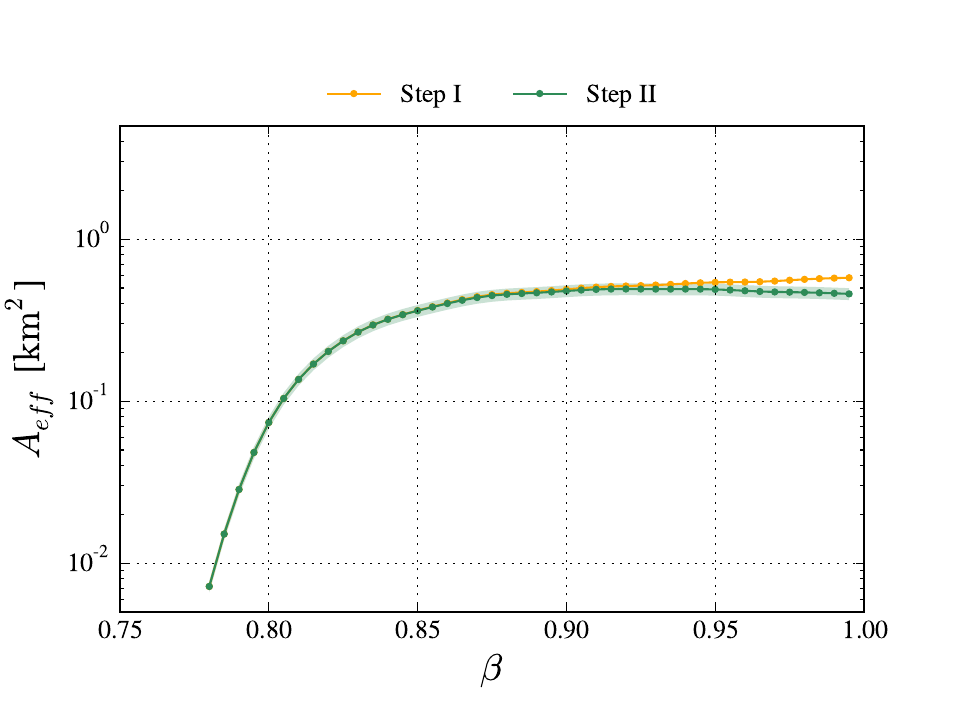}
    \caption{Zenith averaged effective area for monopoles as a function of true particle $\beta$ for the two levels of the analysis. The shaded area in the step II curve represents the systematic uncertainty.}
    \label{fig:Aeff}
\end{figure}

Since the step I analysis targets all neutrino flavors, it does not reject the typical event pattern induced by electron  and tau neutrino charged-current interactions, neither all-flavor neutral current interactions. These events are characterized by the full neutrino energy being deposited within a few meters of the interaction point, producing a spherical light distribution in the detector. A second level of cuts  (denoted as step II in the rest of the Letter) was developed in a blind manner, using only Monte Carlo simulations of signal and background, aimed specifically at selecting monopole tracks, while rejecting the remaining astrophysical neutrino events accepted by step I. A track reconstruction allowing monopole speeds below the speed of light and characterizing the stochastic energy losses along the track was performed at this level~\cite{Aartsen:2013bfa}. Nine variables related to the characteristics of the events were used in a boosted decision tree (BDT) classification of the remaining events.  The variables describe the expected characteristics of monopole events (through-going tracks with a uniform light emission along their path and a given expected total brightness) and allow efficient separation between signal and background. The BDT was trained on a subsample of the simulated signal and background samples to obtain optimal separation between monopoles and astrophysical neutrinos. Optimal separation was defined by minimizing the model rejection potential (MRP)~\cite{Hill:2002nv}, which provides maximum sensitivity to the signal. Figure~\ref{fig:BDT_score} illustrates the separation power of the BDT. The left plot shows the event rate as a function of BDT score for the different particle backgrounds and monopole signal, while the right plot shows the MRP as a function of BDT score. The minimum corresponds to the BDT score that defines the best sensitivity of the analysis, and determines the choice of the optimal BDT cut at a value of 0.047. We assume no uncertainty in this value since the minimum of the score cut distribution is well defined. Note that the BDT score distributions for electron and tau neutrinos are well separated from the distribution for monopoles, while for muon-neutrinos the separation is less pronounced. This reflects the fact that muon tracks can mimic monopole tracks more easily than particle cascades from electron and tau neutrino interactions. The last row of Table~\ref{tab:results} shows the remaining fraction of observed, signal and background events after step II. Figure~\ref{fig:Aeff} shows the effective area of the analysis as a function of monopole speed, averaged in arrival directions over the full hemisphere, after step I cuts and at final analysis level (step II).\\

\paragraph{Systematic uncertainties.---} Uncertainties on the scattering and absorption of Cherenkov photons during their propagation in the ice, as well as of the angular and total sensitivity of the DOMs, limit the sensitivity of IceCube. In order to include these systematic uncertainties in the final result, a study was performed by simulating $10^5$ monopole events for each of ten independent variations of the nominal parameter values used in the standard simulation. Four combinations of the scattering and absorption coefficients of the ice~\cite{Ackermann:2006pva} were tested by shifting their nominal values by $\pm 5\%$ and the DOM total efficiency was shifted by $\pm 10\%$ for all DOMs. Four models of the DOM angular sensitivity that bracket the allowed variations in this quantity were used to evaluate the systematic uncertainty induced by this variable. Each of these variations have been extensively vetted by calibration studies in IceCube and represent the current best knowledge of the detector medium and response parameters. For each parameter variation the monopole events were passed through the full event selection and the effective area was calculated in each case. The variations result in changes of the effective area between +5\% and $-$7\% with respect to the baseline analysis, dominated by the ice absorption. The angular response of the DOMs has a negligible impact in this analysis due to the high brightness of monopole events. The systematic uncertainties from the ice properties, the DOM efficiency and the DOM angular response are assumed independent of each other and are added in quadrature, resulting in a total uncertainty on the sensitivity of 8.4\%. Uncertainties from the intrinsic finite sizes of the Monte Carlo simulated samples used are negligible in this analysis.

An additional source of uncertainty on the magnetic monopole flux limit is the number of expected background events from the astrophysical neutrino flux. The uncertainties in the normalization and spectral index of this flux lead to an uncertainty in the number of predicted background events $n_{\mathrm{bg}}$ of about a factor of 2, $n_{\mathrm{bg}}=0.27^{+0.27}_{-0.14}$. The main contribution to this background arises from astrophysical muon neutrinos ($0.24^{+0.23}_{-0.12}$ events remaining), the other flavors being successfully rejected by the analysis due to their different topology ($0.003^{+0.003}_{- 0.002}$ and $0.021^{+0.003}_{- 0.013}$ electron and tau neutrinos remaining respectively). However, the low statistics of astrophysical neutrinos with energies above 100~TeV allows for alternative fits to the spectrum that result in different expected number of background events. Two additional fits were examined in this analysis: an update to the diffuse high-energy muon-neutrino flux using 10 years of data~\cite{Stettner:2019tok} (yielding a background of $9.4^{+7.3}_{-3.9}$ events after step I and $0.25^{+0.19}_{-0.11}$ events after step II)  and an analysis of high-energy starting events using 7.5 years of data~\cite{IceCube:2020wum} (resulting in $1.1^{+1.7}_{-0.6}$ background events after step I and $0.03^{+0.05}_{-0.02}$ background events after step II). These two assumptions represent extremes of the current allowed range of normalization and spectral index of the astrophysical neutrino flux obtained from IceCube data, assuming a one component power-law fit without cutoff. Each of the evaluated fluxes are compatible with zero events over the full analysis live time within uncertainties. The flux assumed in this work predicts the highest number of events and represents an upper estimation of the astrophysical neutrino background. \\

\paragraph{Results.---}
\begin{table}
\caption{\label{tab:results}
Number of observed events, $n_{\mathrm{obs}}$, at each level of the analysis, along with the corresponding
numbers of expected signal and background (astrophysical neutrino) events, $n_{\mathrm{sg}}$ and $n_{\mathrm{bg}}$, assuming the model fluxes described in the text.}
\begin{ruledtabular}
  \begin{tabular}{r l  r  r r r r c} 
Analysis level  &  ~  &   $n_{\mathrm{obs}}$  &  ~  & $n_{\mathrm{sg}}$  &  ~  &  $n_{\mathrm{bg}}$  & ~ \\ \hline

~ & & & & & & \\[-0.2cm]
Online filter      &  ~        &  $1.63\times 10{^8}$  &  ~  &  178   &  ~  &  371    &  ~  \\ \hline
~ & & & & & & \\[-0.2cm]
 Step I: & & & & & & \\
  Initial off-line cuts &   &  $3.16\times 10^{4}$  &  ~  &  89.9  &  ~  &  57.2   &  ~  \\
          Track quality cut &    &  $8.46\times 10^{3}$ &  ~  &  64.1  &  ~  &  20.4   &  ~  \\
          Down-going cut&  &  3  &  ~  &  35.5  &  ~  &  10.1   &  ~  \\
          IceTop surface veto &   &  3  &  ~  &  35.5  &  ~  &  $10.0^{+10.3}_{-5.1}$  &  ~  \\ \hline
~ & & & & & & \\[-0.2cm]
Step~II  & &     0  &  ~  &  33.2  &  ~  &  $0.27^{+0.27}_{-0.14}$  &  ~  \\

\end{tabular}
\end{ruledtabular} 
\end{table}

\begin{figure}
    \centering
    \includegraphics[width=0.5\textwidth]{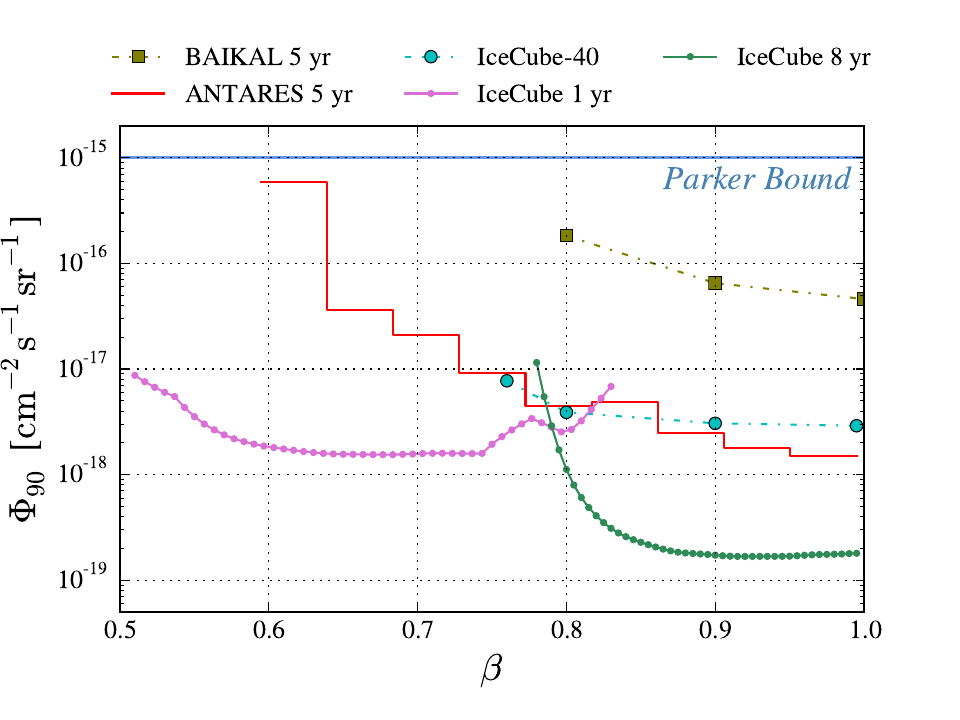}
    \caption{90\% confidence level upper limit on the cosmic flux of relativistic monopoles as a function of true particle $\beta$ obtained in the present analysis assuming zero background (dark green curve). Also included are previous results of IceCube~\cite{Aartsen:2015exf,Aartsen:2014awd}, ANTARES~\cite{Albert:2017fud} and Baikal~\cite{Antipin:2007zz}. The limits are valid for monopoles with the given $\beta$ at the detector. The Parker bound~\cite{Parker:1970xv,Turner:1982ag} is shown as reference.}
    \label{fig:UL_result}
\end{figure}

Table~\ref{tab:results} shows the number of observed events, expected signal events, and background events from the online filter level to the final cut level. The number of expected signal events is given for illustration, assuming a monopole flux at the level of the previous upper limit published in~\cite{Aartsen:2015exf}. The number of background and observed events determine the limit on the cosmic relativistic monopole flux that can be set. Since zero events are observed, consistent with the background expectation, we assume a zero background hypothesis. This assumption produces a conservative upper limit on the monopole flux. The limit has been obtained including systematic uncertainties following the method in Refs.~\cite{Aartsen:2015exf,Conrad:2002kn}, and it is shown as a function of monopole speed in Fig.~\ref{fig:UL_result}, along with results from previous IceCube  analyses and other experiments. Using the background expectation of $0.27^{+0.27}_{-0.14}$ events would result in a more stringent limit by about 10\% due to the underfluctuation of the observation with respect to the expected background. Searches for magnetic monopoles have also been performed by RICE~\cite{Hogan:2008sx}, ANITA~\cite{Detrixhe:2010xi} and Auger~\cite{Aab:2016poe}. These limits are valid in the ultrarelativistic region, $\gamma \sim 10^{7-13}$, which would correspond to a single point at the rightmost end of the $\beta$ axis of Fig.~\ref{fig:UL_result}. The most stringent limit between $\gamma \sim 10^{7}$ and $\gamma \sim 10^{9}$ is provided by the RICE Collaboration, at a level of about $10^{-19} {\textrm{cm}}^{-2} {\textrm{s}}^{-1} {\textrm{sr}}^{-1}$, while the Auger limit reaches down to about $2 \times 10^{-21} {\textrm{cm}}^{-2} {\textrm{s}}^{-1} {\textrm{sr}}^{-1}$ for $\gamma > 10^{11}$. The limit presented in this Letter is the strictest constraint on the flux of magnetic monopoles above $\beta \gtrsim 0.80$ and up to $\beta \sim 0.995$, for monopole masses above $10^{8-10}$~GeV (depending on zenith angle) and complements existing limits on the flux of nonrelativistic and ultrarelativistic monopoles.

\paragraph{Acknowledgments.} 
  The IceCube Collaboration acknowledges the significant contribution to this manuscript from the Uppsala University group.
  We acknowledge the support from the following agencies: 
USA {\textendash} U.S. National Science Foundation-Office of Polar Programs,
U.S. National Science Foundation-Physics Division,
Wisconsin Alumni Research Foundation,
Center for High Throughput Computing (CHTC) at the University of Wisconsin{\textendash}Madison,
Open Science Grid (OSG),
Extreme Science and Engineering Discovery Environment (XSEDE),
Frontera computing project at the Texas Advanced Computing Center,
U.S. Department of Energy-National Energy Research Scientific Computing Center,
Particle astrophysics research computing center at the University of Maryland,
Institute for Cyber-Enabled Research at Michigan State University,
and Astroparticle Physics computational facility at Marquette University;
Belgium {\textendash} Funds for Scientific Research (FRS-FNRS and FWO),
FWO Odysseus and Big Science programmes,
and Belgian Federal Science Policy Office (Belspo);
Germany {\textendash} Bundesministerium f{\"u}r Bildung und Forschung (BMBF),
Deutsche Forschungsgemeinschaft (DFG),
Helmholtz Alliance for Astroparticle Physics (HAP),
Initiative and Networking Fund of the Helmholtz Association,
Deutsches Elektronen Synchrotron (DESY),
and High Performance Computing cluster of the RWTH Aachen;
Sweden {\textendash} Swedish Research Council,
Swedish Polar Research Secretariat,
Swedish National Infrastructure for Computing (SNIC),
and Knut and Alice Wallenberg Foundation;
Australia {\textendash} Australian Research Council;
Canada {\textendash} Natural Sciences and Engineering Research Council of Canada,
Calcul Qu{\'e}bec, Compute Ontario, Canada Foundation for Innovation, WestGrid, and Compute Canada;
Denmark {\textendash} Villum Fonden and Carlsberg Foundation;
New Zealand {\textendash} Marsden Fund;
Japan {\textendash} Japan Society for Promotion of Science (JSPS)
and Institute for Global Prominent Research (IGPR) of Chiba University;
Korea {\textendash} National Research Foundation of Korea (NRF);
Switzerland {\textendash} Swiss National Science Foundation (SNSF);
United Kingdom {\textendash} Department of Physics, University of Oxford.


\end{document}